\documentclass[12pt]{elsarticle}
\usepackage[utf8]{inputenc}
\usepackage{graphicx}
\usepackage{subfigure}
\usepackage{array}
\usepackage{verbatim}

\usepackage{amssymb}
\usepackage{amsthm}
\usepackage{url}
\usepackage{amsmath}
\usepackage{subfigure}
\usepackage{textcomp}
\usepackage{xcolor}

\graphicspath{{figs/}}

\journal{BioSystems}

\begin{document}

\begin{frontmatter}

\title{Mining logical circuits in fungi}
\author{Nic Roberts}
\author{Andrew Adamatzky}

\address{Unconventional Computing Laboratory, UWE, Bristol, UK}

\begin{abstract}
\noindent
Living substrates are capable for nontrivial mappings of electrical signals due to the substrate nonlinear electrical characteristics. This property can be used to realise Boolean functions. Input logical values are represented by amplitude or frequency of electrical stimuli. Output logical values are decoded from electrical responses of living substrates. We demonstrate how logical circuits can be implemented in mycelium bound composites. The mycelium bound composites (fungal materials) are getting growing recognition as building, packaging, decoration and clothing materials. Presently the fungal materials are passive. To make the fungal materials adaptive, i.e. sensing and computing, we should embed logical circuits into them. We demonstrate experimental laboratory prototypes of many-input Boolean functions implemented in fungal materials from oyster fungi \emph{P. ostreatus}. We characterise complexity of the functions discovered via complexity of the space-time configurations of one-dimensional cellular automata governed by the functions. We show that the mycelium bound composites can implement representative functions from all classes of cellular automata complexity including the computationally universal. The results presented will make an impact in the field of unconventional computing, experimental demonstration of purposeful computing with fungi, and in the field of intelligent materials, as the prototypes of computing mycelium bound composites.
\end{abstract}


\end{frontmatter}

\section{Introduction}

The fungi are one of the largest, the oldest, most adaptive and widely distributed group of organisms~\cite{carlile2001fungi}. Smallest fungi are single cells. The largest mycelium spreads in hectares~\cite{smith1992fungus}. When growing in a bulk medium of wood or plant shavings fungi bind the medium in a solid monolith with outstanding mechanical properties. The mycelium bound composites are seen as future environmentally sustainable growing biomaterials~\cite{karana2018material,jones2020engineered,cerimi2019fungi,adamatzky5adaptive}. They are already used in acoustic~\cite{pelletier2013evaluation,elsacker2020comprehensive,robertson2020fungal} and thermal~\cite{yang2017physical,xing2018growing,girometta2019physico,dias2021investigation,wang2016experimental,cardenas2020thermal} insulation panels and cladding, materials for packaging~\cite{holt2012fungal,sivaprasad2021development,mojumdar2021mushroom} and wearables~\cite{adamatzky2021reactive,silverman2020development,karana2018material,appels2020use,jones2020leather}. The currently used fungal materials are passive and inert because the fungi in the composites are dead and treated to prevent decay. To make the fungal materials adaptive and intelligent we must either (1)~leave part of the fungal materials alive, or (2)~dope the materials with functional nanoparticles and polymers. In the present paper we explore the first option of sensing and computing with living mycelium. 

Fungal colonies are characterised by rich typology  of mycelium networks~\cite{hitchcock1996image,giovannetti2004patterns,fricker2007network,fricker2017mycelium,islam2017morphology} in some cases similar to fractal structures~\cite{obert1990microbial,patankar1993fractal,bolton1993characterization,mihail1995fractal,boddy1999fractal,papagianni2006quantification}. Rich morphological features might imply rich computational abilities and thus worth to analyse from a realising Boolean functions point of view. To implement logical functions we adopted a theoretical approach developed in  \cite{adamatzky2020boolean,siccardi2015actin}. The technique is based on selecting a pair of input sites, applying all possible combinations of inputs, where logical values are represented by electrical characteristics of input signals, to the sites and recording outputs, represented by electrical responses of the substrate, on a set of the selected output sites. The approach belong to the family of reservoir computing~\cite{verstraeten2007experimental,lukovsevivcius2009reservoir,dale2017reservoir,konkoli2018reservoir,dale2019substrate} and   \emph{in materio} computing~\cite{miller2002evolution,miller2014evolution,stepney2019co,miller2018materio,miller2019alchemy} techniques of analysing computational properties of physical and biological substrates. 

The paper is structured as follows. First, the experimental setup will be described, then the procedure for data gathering and analysis will be outlined.

\section{Methods}
\label{methods}

\begin{figure}[!tbp]
    \centering
    \includegraphics[scale=0.75]{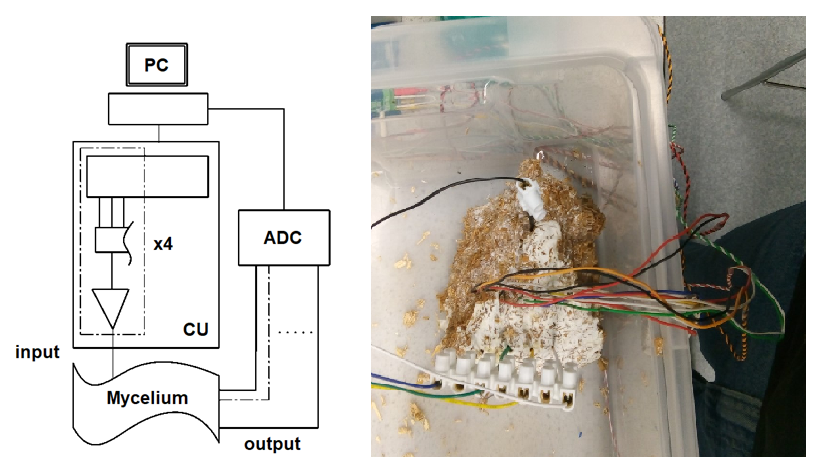}
    \caption{Left: Schematic of the mycelium communications system; PC –- laptop for generating sequences; CU -- control unit, dashed section is a breakdown of a single channel; ADC –- analogue to digital converter. Right: experimental set up.}
    \label{fig:schematic}
\end{figure}

A hemp shavings substrate was colonised by the mycelium of the grey oyster fungi, \emph{P. ostreatus} (Ann Miller's Speciality Mushrooms Ltd, UK). Recordings were carried out in a stable indoor environment with the temperature remaining stable at $22 \pm 0.5$\textdegree and relative humidity of air $40 \pm 5$\%. The humidity of the substrate colonised by fungi was kept at c. 70-80\%. 

Hardware was developed that was capable of sending sequences of 4 bit strings to a mycelium substrate. The strings were encoded as step voltage inputs where -5~V denoted a logical 0 and 5~V a logical 1. The hardware was based around an Arduino Mega 2560 (Elegoo, China) and a series of programmable signal generators, AD9833 (Analog, USA). 
The 4 input electrodes were 1~mm diameter platinum rods inserted to a depth of 50~mm in the substrate in a straight line with a separation of 20~mm. Data acquisition (DAQ) probes were placed in a parallel line 50~mm away separated by 10~mm. The electron sink and source was placed 50~mm on from DAQ probes. There were 7 DAQ differential inputs from the mycelium substrate to a Pico 24 (Pico Technology, UK) analogue-to-digital converter (ADC), the 8th channel was used to pass a pulse to the ADC on every input state change, see Fig.~\ref{fig:schematic} for a schematic of the apparatus. The substrate and probes were placed in a semi-sealed container. After each experimental repeat the substrate was sprayed with water, left for an hour and then the next repeat was conducted. There were a total of 14 repeats.

\begin{figure}[!tbp]
    \centering
    \includegraphics[scale=0.5]{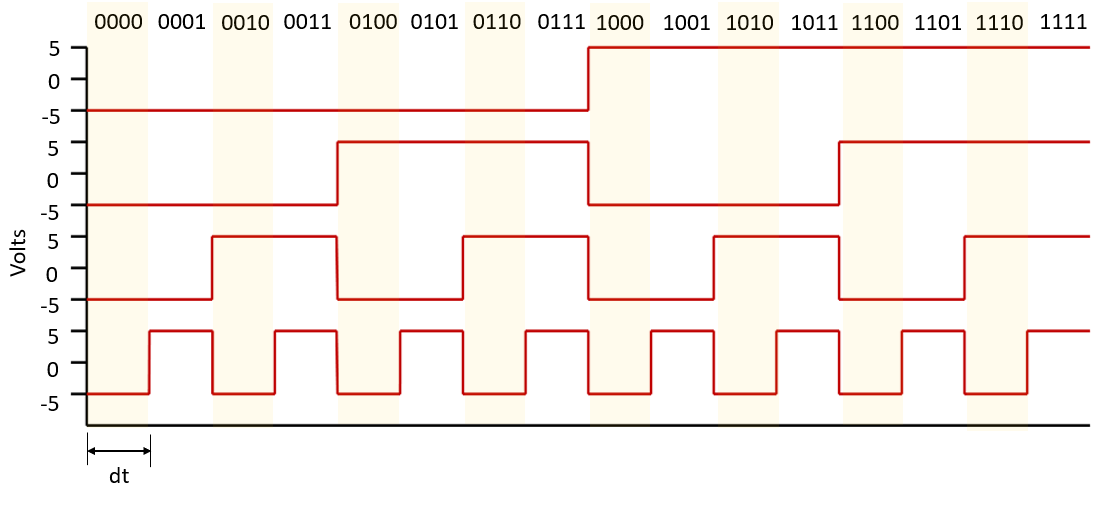}
    \caption{Timing diagram and associated Boolean strings for four inputs into the mycelium substrate, time step is  one hour.}
    \label{fig:timing}
\end{figure}

A sequence of 4 bit strings counting up from binary \textit{0000} to \textit{1111}, with a state change every hour, were passed into the substrate, see Fig.~\ref{fig:timing} for timing details. In all 14 repeats of the experiment were done on the same substrate to capture changes in structure of the growing mycelium. 
Samples from 7 channels were taken at 1~Hz over the whole duration of a given experimental run. Peaks for each channel were located for a set of 32 thresholds, from 20~mV to 175~mV with step 5~mV, for each input state, \textit{0000 to 1111}.

Boolean strings were extracted from the data, where a logic ‘1’ was noted for a channel if it had a peak outside the threshold band for a particular state else, a value of ‘0’ was recorded, the polarity of the peak was not considered.

The strings for each experimental repeat were stored in their respective Boolean table. To extract state graphs, a state/node was defined as the string of output values from each channel at each input state, transitions/edges were defined as a change in input state. This led to a total of 448 state graphs.
The sum of products (SOP) Boolean functions were calculated for each output channel. For each repeat there were 7 channels and 32 thresholds giving total of 3136 individual truth tables.

\begin{figure}[!tbp]
    \centering
    \subfigure[]{\includegraphics[width=\textwidth]{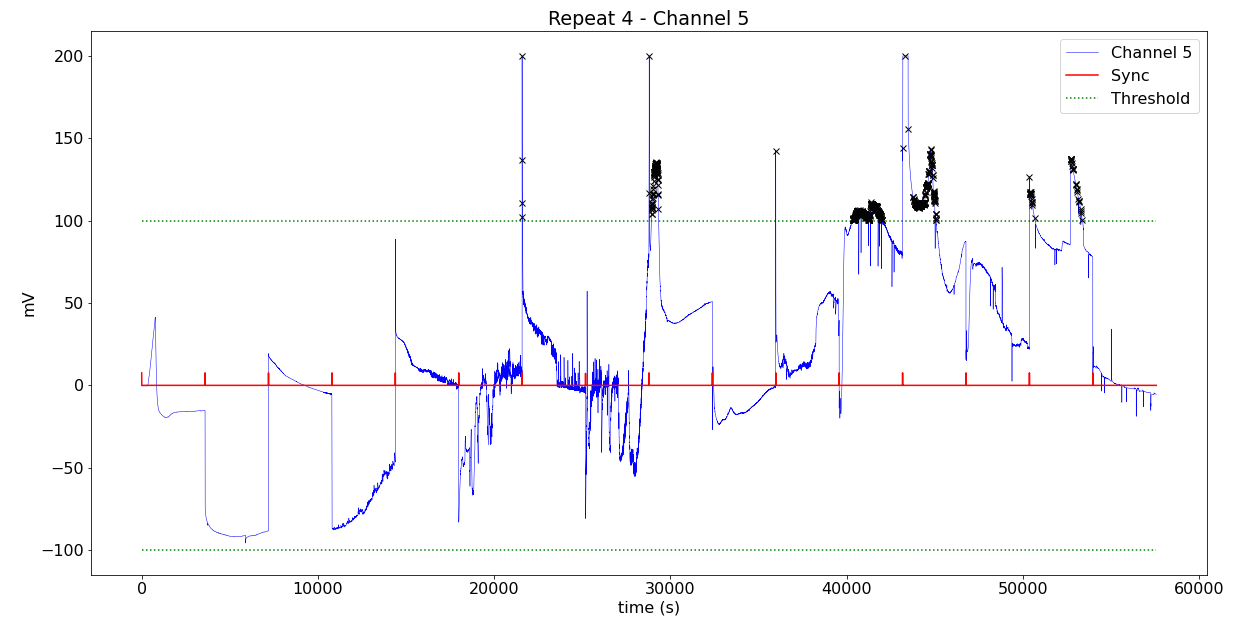}}
    \subfigure[]{\includegraphics[width=0.6\textwidth]{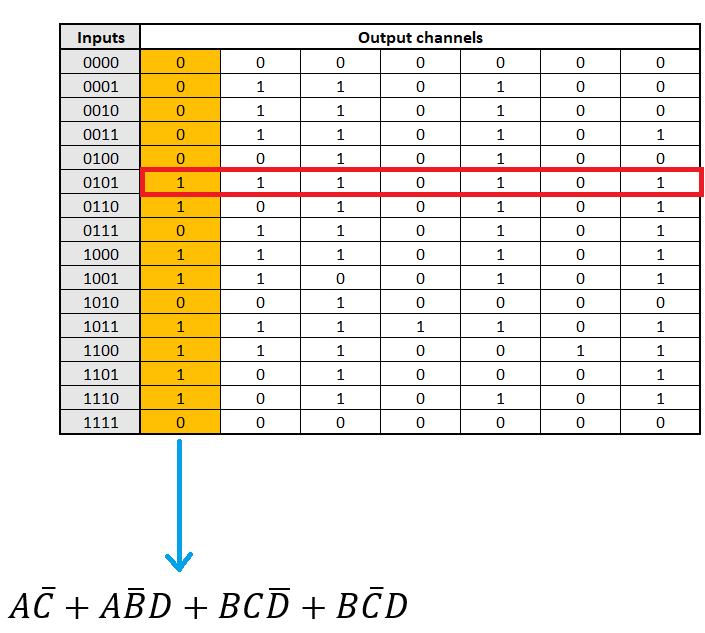}}
    \caption{Workflow example. 
    (a)~The measurements taken by channel 5 of the DAQ in blue, the synchronisation signal is shown red which marks the state change, threshold band shown in green, peaks outside this band are highlighted with `x' marker. 
    (b)~The truth and the function extracted. 
    }
    \label{fig:sample_data}
\end{figure}

See Fig.~\ref{fig:sample_data} for  SOP extraction. If a peak is found in Fig.~\ref{fig:sample_data}a during an input state then this is considered a logical 1, highlighted in yellow in table Fig.~\ref{fig:sample_data}b are the thresholded values for channel 5, the resulting truth table is then reduced to a sum products shown below the table.

\section{Results}

\begin{figure}[!tbp]
    \centering
    \includegraphics[width=\textwidth]{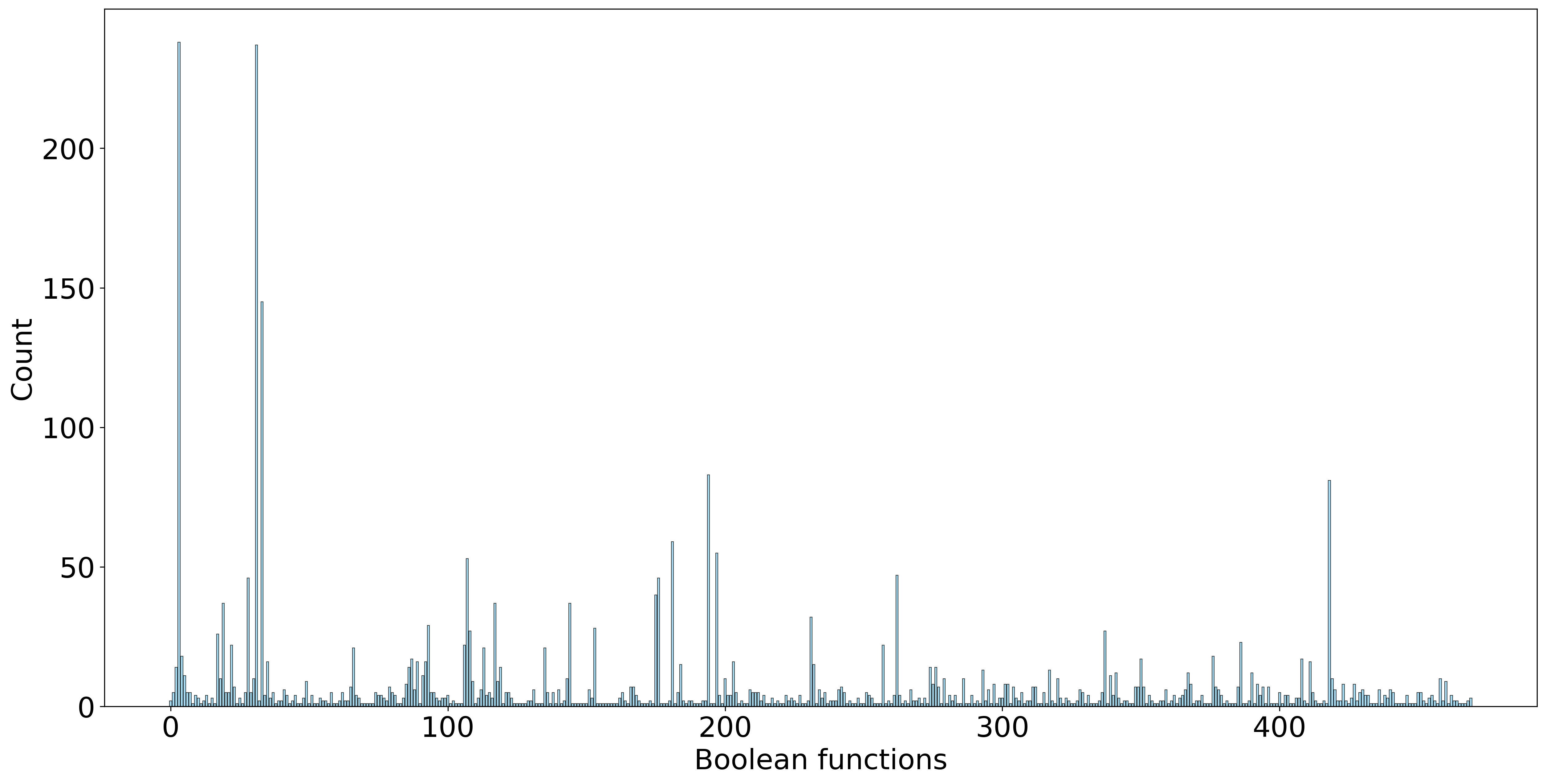}
    \caption{Counts of realised Boolean functions discovered in laboratory experiments. Horizontal axis is a decimal representation of functions. Vertical axis is a number of functions discovered in experiments.}
    \label{fig:Bool_func_count}
\end{figure}

\begin{table}[!tbp]
 \caption{Top 16 highest occurring Boolean functions.}
\centering
  \begin{tabular}{ |p{1cm}|p{1cm}|p{12cm}|  }
   \hline
   Count & & Boolean function\\
   \hline
    145 & $F_1$ &$\overline{A}+\overline{B}+\overline{C}+\overline{D}$ ({\sc nand})\\
    83 & $F_2$ & $A\overline{B}+A\overline{C}+A\overline{D}+\overline{A}B+B\overline{C}+B\overline{D}+\overline{A}C+\overline{B}C+C\overline{D}+\overline{A}D+\overline{B}D+\overline{C}D$\\
    81 & $F_3$ & $AC\overline{D}+\overline{A}B\overline{C}+\overline{A}\overline{B}C+\overline{A}\overline{B}D$\\
    59 & $F_4$ & $A\overline{C}+A\overline{D}+\overline{A}C+C\overline{D}+\overline{A}D+\overline{B}D+\overline{C}D$\\
    55 & $F_5$ & $\overline{A}B+C\overline{D}+\overline{A}D$\\
    53 & $F_6$ & $A\overline{B}CD$\\
    47 & $F_7$ & $B\overline{D}+C\overline{D}+\overline{A}D+\overline{B}\overline{C}D$\\
    46 & $F_8$ & $AB\overline{C}\overline{D}$\\
    46 & $F_9$ & $A+B+C+D$ ({\sc or})\\
    40 & $F_{10}$ & $A\overline{B}+A\overline{D}+\overline{A}B+B\overline{D}+\overline{A}D+\overline{B}D+\overline{C}D$\\
    37 & $F_{11}$ & $A\overline{B}\overline{C}\overline{D}$\\
    37 & $F_{12}$& $A\overline{D}+\overline{A}B+B\overline{C}+\overline{A}D+\overline{B}CD$\\
    37 &$F_{13}$  & $A\overline{B}+A\overline{C}+A\overline{D}+\overline{A}D+\overline{B}D+\overline{C}D\overline{A}BC+BC\overline{D}$\\
    32 & $F_{14}$ & $A\overline{D}+\overline{A}B+B\overline{D}+\overline{A}C+C\overline{D}+\overline{A}D+A\overline{B}\overline{C}+\overline{B}\overline{C}D$\\
    29 & $F_{15}$ & $\overline{C}+A\overline{B}+A\overline{D}+\overline{A}B+B\overline{D}\overline{A}D+\overline{B}D$\\
    28 & $F_{16}$ & $\overline{A}B+\overline{A}C+\overline{B}D+BC\overline{D}+A\overline{B}\overline{C}$\\
   \hline
  \end{tabular}
 \label{table:1}
\end{table}

We have discovered total of 3136 4-inputs-1-output Boolean functions. 470 unique functions are presented in Supplementary Materials. Figure~\ref{fig:Bool_func_count} shows the Boolean function distribution. The two peak values were logical {\sc False}, $n=238$, and logical {\sc True}, $n=237$. The highest occurring non-trivial gate was $\overline{A}+\overline{B}+\overline{C}+\overline{D}$, $n=145$. The top 16 occurring non-trivial Boolean functions are listed in Tab.~\ref{table:1}. The only single gate functions found were for {\sc nand} ($\overline{A}+\overline{B}+\overline{C}+\overline{D}$), $n=145$, {\sc or} ($A+B+C+D$), $n=46$, and {\sc and} ($ABCD$), $n=8$.


\begin{figure}[!tbp]
    \centering
    \subfigure[$F_2$]{\includegraphics[width=0.24\textwidth]{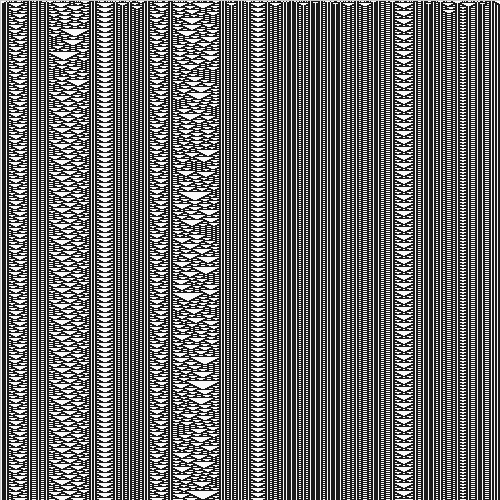}}
     \subfigure[$F_3$]{\includegraphics[width=0.24\textwidth]{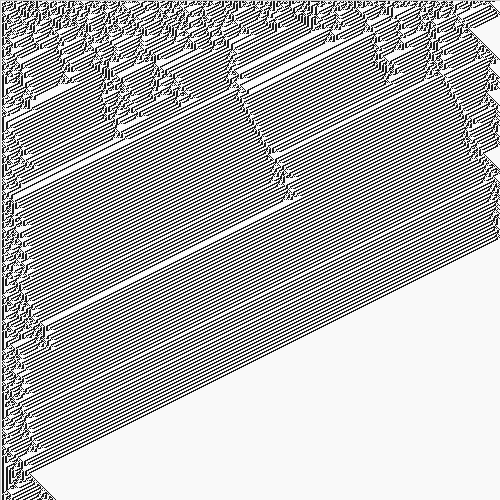}}
     \subfigure[$F_4$]{\includegraphics[width=0.24\textwidth]{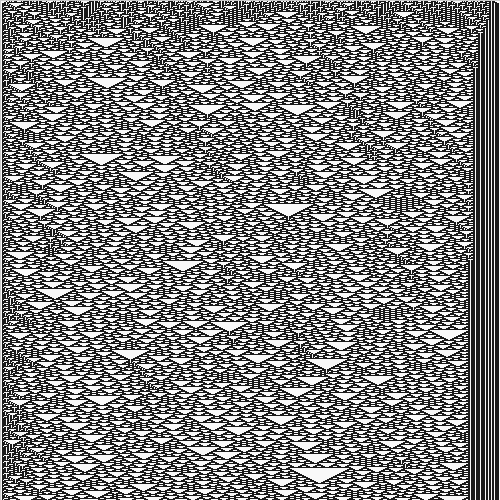}}
         \subfigure[$F_5$]{\includegraphics[width=0.24\textwidth]{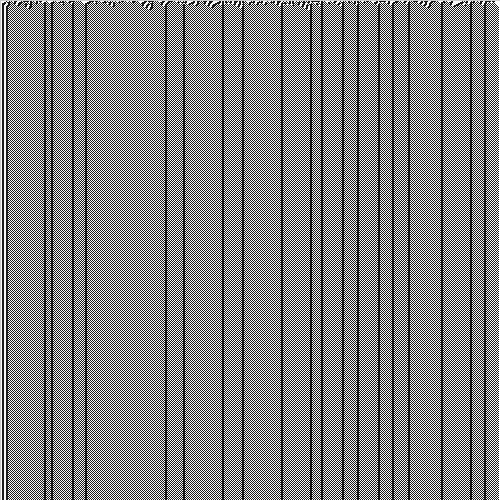}}
         \subfigure[$F_7$]{\includegraphics[width=0.24\textwidth]{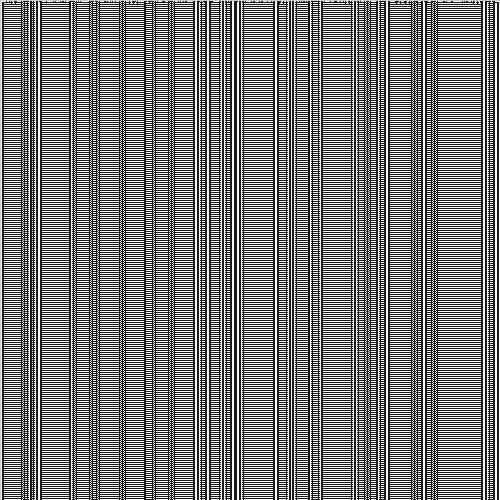}}
\subfigure[$F_{10}$]{\includegraphics[width=0.24\textwidth]{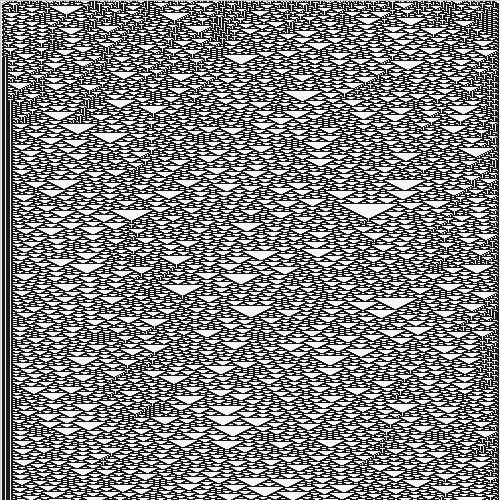}}
    \subfigure[$F_{11}$]{\includegraphics[width=0.24\textwidth]{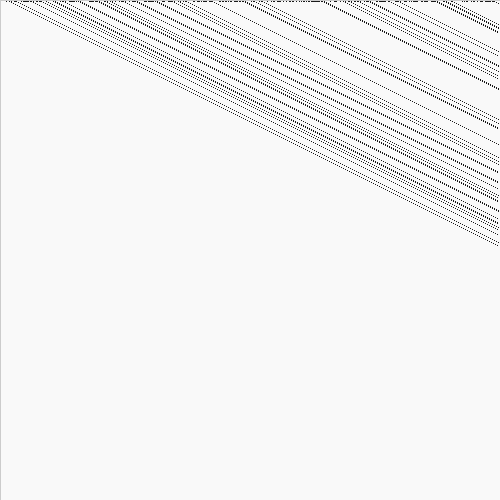}}
\subfigure[$F_{12}$]{\includegraphics[width=0.24\textwidth]{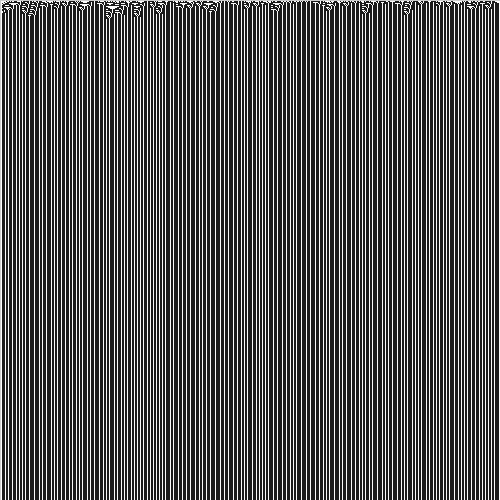}}
\subfigure[$F_{13}$]{\includegraphics[width=0.24\textwidth]{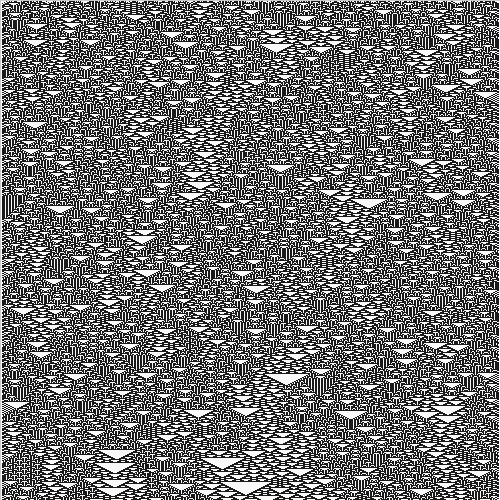}}
\subfigure[$F_{14}$]{\includegraphics[width=0.24\textwidth]{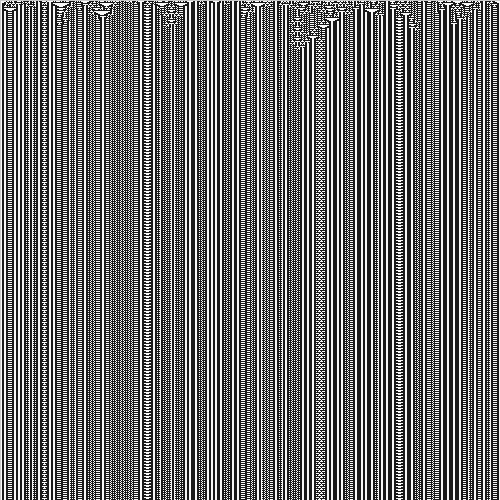}}
\subfigure[$F_{15}$]{\includegraphics[width=0.24\textwidth]{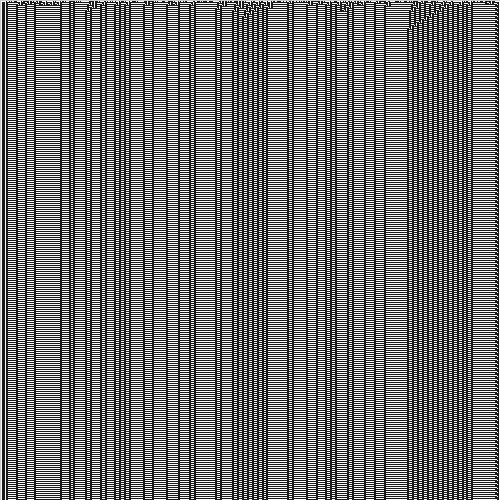}}
\subfigure[$F_{16}$]{\includegraphics[width=0.24\textwidth]{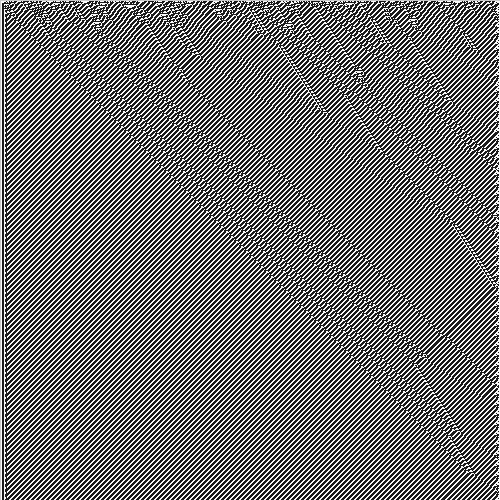}}
    \caption{Space-time configurations of one-dimensional cellular automata governed by functions from Tab.~\ref{table:1}. An automaton has 500 cells and evolves for 500 iterations. Initial configurations has a random uniform distribution of cells in state `1' where each cell takes the state `1' with a probability $\frac{1}{2}$.}
    \label{fig:spacetime}
\end{figure}

\begin{figure}[!tbp]
    \centering
    \includegraphics[width=0.7\textwidth]{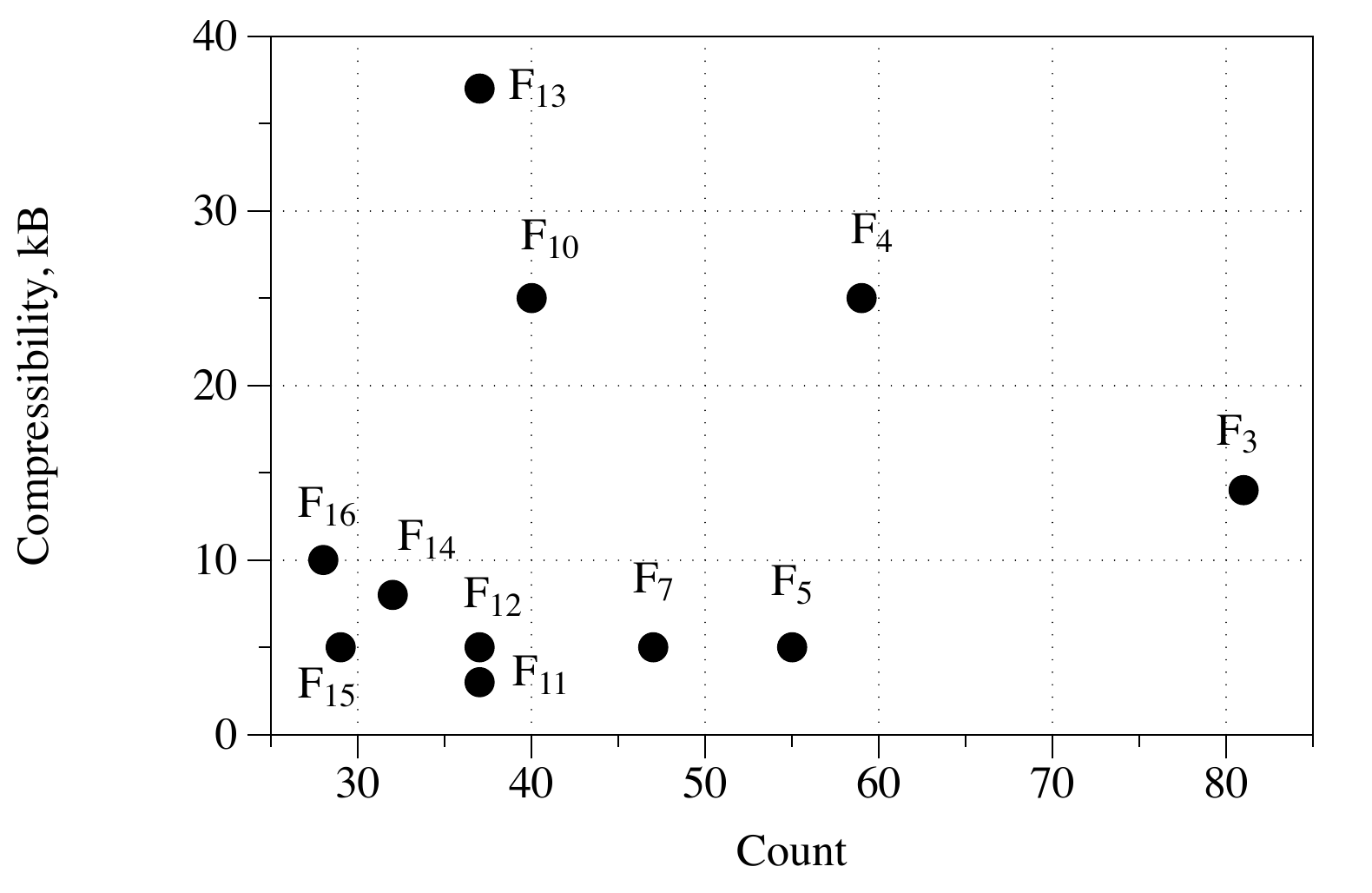}
    \caption{Frequency of functions from Tab.~\ref{table:1} versus $LZ$ complexity, measured via compressibility of the space-time configurations of cellular automata governed by the functions. Functions $F_1$, $F_6$, $F_8$ and $F_9$ are not displayed because their $LZ$ is near zero.}
    \label{fig:countvsLZ}
\end{figure}

Let us discuss complexity of the functions discovered (Tab.~\ref{table:1}) via complexity of the space-time configurations of one-dimensional cellular automata governed by the functions. We consider an array $Z$ of finite state machines, called cells, where every cell takes states `0' or `1' and updates its state depending on the states of its four immediate neighbours. All cells update their states by the same rule and in discrete time. For example, a cell with index $i$, $x_i \in Z$, updates its state at time $t$ as a function of states of its four neighbours: $x^{t+1}=f(x_{i-2}^t, x_{i-1}^t, x_{i+1}^t, x_{i+2}^t)$.  To map functions from Tab.~\ref{table:1} to the rules governing the cellular automata we assume that $A$ corresponds to $x_{i-2}^t$, $B$ to $x_{i-1}^t$, $C$ to  $x_{i+1}^t$ and $D$ to $x_{i+1}^t$. For example, a cell $x_i$ of cellular automaton governed by the function $F_5$ (Tab.~\ref{table:1}) updates its state as $x^{t+1}=
\overline{x_{i-2}}x_{i-1}+x_{i+1}\overline{x_{i+2}}+\overline{x_{i-2}}x_{i+2}$.

Automaton governed by $F_1$, $F_6$, $F_8$ fall into absorbing state where all cells are in state `0'.  The automaton governed by rule $F_9$ falls into the state where all cells are in state `1'. Space-time configurations, random initial conditions and absorbing boundaries, of automata governed by other rules are shown in Fig.~\ref{fig:spacetime}.  We characterise a complexity of the space-time patterns via Lempel-Ziv complexity (compressibility) $LZ$. The $LZ$ complexity is evaluated by a size of concentration profiles saved as PNG files  of the configurations. This is sufficient because the 'deflation' algorithm used in PNG lossless compression~\cite{roelofs1999png,howard1993design, deutsch1996zlib}  is a variation of the classical Lempel--Ziv 1977 algorithm~\cite{ziv1977universal}.
The frequency of the functions occurrence in the experimental circuit mining versus $LZ$ complexity of the functions is shown in Fig.~\ref{fig:countvsLZ}. We can see that there is no correlation between  how often a function can be found and how complexity the function is. Thus, e.g. the function $F_{13}$ (Tab.~\ref{table:1}) generates most complex space-time configuration (Fig.~\ref{fig:spacetime}i) yet it is in the mid-range of the frequency of experimental occurrence. The less complex functions $F_5$, $F_7$, $F_{12}$, $F_{15}$ span the interval [29,55] counts of occurrences in experimental laboratory mining. 

Let us consider positions of the functions Tab.~\ref{table:1} in the Wolfram classification~\cite{wolfram1983statistical} of cellular automaton behaviour. 
Functions $F_1$, $F_6$, $F_8$, $F_9$ and $F_11$ belong to the class I, the class of automata exhibiting a dull dynamics and evolving to a stable state where all cells are in the same state.  Functions $F_2$, $F_7$, $F_{12}$, $F_{14}$, $F_{15}$ belong to the class II: the automata fall into global cells do not update their state or update them cyclically from `0' to `1'. Functions $F_4$, $F_{10}$ and $F_{13}$ belong to class III: the space-time dynamics is characterised by quasi-random behaviour and difficult predictability of the successions of the global states. These functions generate the most complex, as evaluated by $LZ$ measure, space-time configurations. Function $F_2$ shows an interesting example of the function belonging to classes II and III. Two functions $F_3$ and $F_{16}$ belong to class IV: the space-time dynamics of automata show gliders (compact patterns translating in space) with non-trivial interactions between the gliders. The automata governed by rules $F_3$ and $F_{16}$ are computationally universal, because it is possible to implement an arbitrary logical circuit via collisions between the gliders, see e.g. ~\cite{martinez2006phenomenology,martinez2011cellular}. 

\section{Discussion}

Mycelium bound composites transform electrical signals in a non-linear manner due to mem-fractive and capacitive properties of the fungal tissue~\cite{beasley2020mem}. Whilst exact biophysical mechanisms of the signal transformation by the mycelium remain unknown we can explore the non-linear properties of this living substrate to implement logical circuits. In experimental laboratory studies we demonstrated that mycelium bound composites implement a wide range of Boolean circuits. Analyses of the functions extracted in terms of space-time dynamics of cellular automata helped us to order the functions in several classes of complexity and pinpoint the functions supporting a universal computation. The first ever prototype of the fungal reservoir computer, presented in the paper, demonstrates that a computation can be embedded into living materials. The research presented also pinpointed a high degree of variability in the logical circuits implemented by the fungi. This is because the live mycelium remain in the continuous process of growth and reconfiguration. To decrease the variability of the results we could consider to functionalise the mycelium networks with semi-conductive particles and polymers and allow the mycelium to dry. The resulting networks will have a permanent structure which will guarantee repeatability of the experimental circuits discovered. This will be a topic of our future studies.

\section*{Acknowledgement}

This project has received funding from the European Union's Horizon 2020 research and innovation programme FET OPEN ``Challenging current thinking'' under grant agreement No 858132.

\section*{Supplementary materials}

4-inputs-1-output logical functions discovered in experiments with mycelium bound composites.

\vspace{5mm}

\noindent
$
(A  \overline{D}) + (C  \overline{B}) + (D  \overline{A}) + (\overline{A}  \overline{B}) + (B  D  \overline{C})\\
(A  \overline{D}) + (D  \overline{A}) + (B  D  \overline{C})\\
(A  D  \overline{C}) + (B  D  \overline{A}) + (A  C  \overline{B}  \overline{D})\\
(B  D) + (A  B  \overline{C}) + (A  C  \overline{B})\\
(A  \overline{D}) + (B  D  \overline{A})\\
(B  C) + (B  D) + (C  D) + (A  \overline{D}) + (D  \overline{A})\\
(A  B  \overline{C}) + (A  C  D  \overline{B})\\
(A  D  \overline{B}) + (D  \overline{B}  \overline{C}) + (B  C  D  \overline{A}) + (A  B  \overline{C}  \overline{D})\\
(B  \overline{D}) + (D  \overline{A}) + (A  C  \overline{D})\\
A + B + \overline{D}\\
(B  C) + (B  D) + (A  C  D) + (A  \overline{C}  \overline{D})\\
(A  \overline{D}) + (C  D  \overline{A})\\
(A  D  \overline{B}) + (A  \overline{C}  \overline{D}) + (D  \overline{B}  \overline{C}) + (B  C  D  \overline{A})\\
(B  C  \overline{A}) + (A  B  D  \overline{C}) + (A  C  D  \overline{B}) + (A  \overline{B}  \overline{C}  \overline{D})\\
A  D  \overline{C}\\
(B  C) + (C  D) + (A  \overline{D}) + (D  \overline{A})\\
\overline{C} + (A  \overline{B}) + (B  \overline{A}) + (D  \overline{A})\\
(A  B  \overline{C}) + (A  D  \overline{B}) + (B  D  \overline{A}) + (C  D  \overline{A})\\
(A  B  \overline{C}) + (A  C  \overline{D})\\
(A  B  C  \overline{D}) + (B  C  D  \overline{A}) + (A  \overline{B}  \overline{C}  \overline{D})\\
(A  C  \overline{B}) + (B  D  \overline{A}) + (A  \overline{B}  \overline{D})\\
(B  \overline{A}) + (C  \overline{A}) + (D  \overline{B}) + (B  C  \overline{D}) + (A  \overline{B}  \overline{C})\\
(A  B  D) + (A  C  D)\\
(A  B  \overline{C}) + (A  C  \overline{B}) + (B  C  \overline{D}) + (B  D  \overline{C}) + (C  D  \overline{B})\\
A  B  D  \overline{C}\\
(A  \overline{B}) + (B  \overline{A}) + (D  \overline{A}) + (B  C  \overline{D})\\
(A  \overline{D}) + (A  B  \overline{C}) + (A  C  \overline{B}) + (B  C  \overline{A}) + (B  C  \overline{D}) + (B  D  \overline{A}) + (B  D  \overline{C}) + (D  \overline{A}  \overline{C}) + (\overline{A}  \overline{B}  \overline{C}) + (\overline{B}  \overline{C}  \overline{D})\\
(A  B  \overline{C}) + (A  B  \overline{D}) + (A  C  \overline{B}) + (A  C  \overline{D}) + (A  D  \overline{B}) + (A  D  \overline{C}) + (B  D  \overline{A}) + (B  D  \overline{C}) + (C  D  \overline{A}) + (C  D  \overline{B})\\
(A  B  \overline{D}) + (A  \overline{C}  \overline{D}) + (A  C  D  \overline{B})\\
(B  \overline{A}) + (C  \overline{A}) + (D  \overline{A}) + (B  C  \overline{D}) + (C  D  \overline{B})\\
(D  \overline{A}) + (D  \overline{B}) + (D  \overline{C}) + (A  B  \overline{C}) + (A  B  \overline{D}) + (A  C  \overline{B}) + (A  C  \overline{D}) + (B  C  \overline{A}) + (B  C  \overline{D})\\
(B  \overline{A}) + (B  \overline{C}) + (B  \overline{D}) + (A  \overline{C}  \overline{D}) + (C  \overline{A}  \overline{D}) + (D  \overline{A}  \overline{C}) + (A  C  D  \overline{B})\\
(A  \overline{B}  \overline{C}  \overline{D}) + (D  \overline{A}  \overline{B}  \overline{C})\\
(A  \overline{D}) + (D  \overline{A}) + (D  \overline{B}) + (D  \overline{C})\\
(D  \overline{A}) + (D  \overline{C}) + (B  C  \overline{A})\\
(A  \overline{D}) + (D  \overline{A}) + (D  \overline{B}  \overline{C})\\
(A  \overline{D}) + (D  \overline{A}) + (A  B  \overline{C}) + (A  C  \overline{B}) + (B  C  \overline{A}) + (B  C  \overline{D}) + (B  D  \overline{C}) + (C  D  \overline{B}) + (\overline{A}  \overline{B}  \overline{C}) + (\overline{B}  \overline{C}  \overline{D})\\
(A  B  \overline{D}) + (A  C  \overline{B}) + (C  D  \overline{A}) + (D  \overline{A}  \overline{B})\\
(A  \overline{D}) + (B  C  \overline{A}) + (D  \overline{A}  \overline{B}  \overline{C})\\
(B  C  \overline{A}) + (B  D  \overline{C}) + (A  C  D  \overline{B}) + (\overline{B}  \overline{C}  \overline{D})\\
(A  \overline{D}) + (C  \overline{B}) + (D  \overline{A}) + (D  \overline{C})\\
(A  D  \overline{C}) + (B  D  \overline{A}) + (C  D  \overline{A})\\
(A  \overline{C}) + (B  C  \overline{D}) + (B  D  \overline{A}) + (C  D  \overline{B})\\
(A  \overline{B}) + (A  \overline{C}) + (D  \overline{A})\\
(A  \overline{B}) + (C  \overline{A}) + (D  \overline{C})\\
(D  \overline{A}) + (D  \overline{B}) + (A  B  \overline{C}) + (B  C  \overline{A})\\
(A  B  \overline{D}) + (A  C  \overline{D})\\
(A  B  C  \overline{D}) + (A  C  D  \overline{B}) + (B  C  D  \overline{A})\\
(B  C) + (A  \overline{D}) + (D  \overline{A}) + (D  \overline{B})\\
(A  C  \overline{B}) + (B  C  \overline{A}) + (B  D  \overline{C}) + (A  \overline{C}  \overline{D})\\
(A  B  \overline{C}) + (A  \overline{B}  \overline{D}) + (B  C  D  \overline{A})\\
(A  \overline{D}) + (B  C  \overline{D}) + (D  \overline{A}  \overline{B}) + (D  \overline{A}  \overline{C})\\
\overline{A} + \overline{B} + (C  \overline{D}) + (D  \overline{C})\\
(B  \overline{A}  \overline{C}) + (B  \overline{A}  \overline{D}) + (B  \overline{C}  \overline{D})\\
(A  \overline{C}  \overline{D}) + (A  C  D  \overline{B}) + (B  C  \overline{A}  \overline{D})\\
(A  B  \overline{C}) + (A  C  D  \overline{B}) + (B  C  \overline{A}  \overline{D})\\
(A  B  \overline{C}) + (A  C  \overline{B}) + (A  D  \overline{B}) + (B  C  \overline{A}) + (\overline{A}  \overline{B}  \overline{C}  \overline{D})\\
(B  \overline{A}) + (C  \overline{A}  \overline{D}) + (A  C  D  \overline{B})\\
(A  \overline{C}  \overline{D}) + (A  C  D  \overline{B})\\
(A  B  \overline{C}) + (A  C  D  \overline{B}) + (B  C  D  \overline{A})\\
(A  D  \overline{C}) + (B  D  \overline{A})\\
\overline{A} + \overline{D} + (B  \overline{C}) + (C  \overline{B})\\
(D  \overline{A}) + (D  \overline{B}) + (D  \overline{C}) + (A  C  \overline{B}) + (A  C  \overline{D}) + (B  C  \overline{A}) + (B  C  \overline{D})\\
(A  \overline{D}) + (D  \overline{A}) + (B  C  \overline{A}) + (B  D  \overline{C}) + (C  D  \overline{B})\\
\overline{C} + (B  D) + (A  \overline{B}) + (B  \overline{A})\\
(B  C) + (B  D) + (C  D) + (D  \overline{A}) + (A  \overline{C}  \overline{D})\\
(A  \overline{D}) + (D  \overline{A}) + (D  \overline{C})\\
(A  \overline{D}) + (B  \overline{C}) + (C  \overline{A}) + (C  \overline{B}) + (D  \overline{A})\\
(A  \overline{C}) + (D  \overline{C}) + (A  D  \overline{B}) + (B  C  \overline{D})\\
B + \overline{A} + \overline{D}\\
(A  C  \overline{D}) + (B  D  \overline{A})\\
\overline{A} + \overline{B} + \overline{D}\\
(A  \overline{D}) + (B  \overline{A}) + (B  \overline{C}) + (C  D  \overline{B})\\
(D  \overline{A}) + (D  \overline{B}) + (A  B  \overline{C})\\
(A  \overline{D}) + (B  \overline{A}) + (B  \overline{C}) + (B  \overline{D}) + (D  \overline{A}) + (\overline{A}  \overline{C}) + (\overline{C}  \overline{D}) + (A  C  \overline{B}) + (C  D  \overline{B})\\
(D  \overline{A}) + (D  \overline{C}) + (A  C  \overline{B}) + (B  C  \overline{A})\\
(A  C  D  \overline{B}) + (A  B  \overline{C}  \overline{D})\\
(A  B  D  \overline{C}) + (A  \overline{B}  \overline{C}  \overline{D}) + (D  \overline{A}  \overline{B}  \overline{C})\\
(A  \overline{D}) + (D  \overline{A}) + (D  \overline{B}) + (D  \overline{C}) + (\overline{B}  \overline{C})\\
(A  \overline{B}) + (A  \overline{C}) + (B  C  \overline{A}) + (B  D  \overline{A}) + (\overline{B}  \overline{C}  \overline{D})\\
(A  \overline{B}  \overline{D}) + (A  \overline{C}  \overline{D})\\
(A  \overline{B}) + (A  \overline{C}) + (A  \overline{D}) + (B  C  \overline{A}) + (B  C  \overline{D}) + (B  D  \overline{A}) + (B  D  \overline{C}) + (C  D  \overline{A}) + (C  D  \overline{B}) + (\overline{B}  \overline{C}  \overline{D})\\
(B  \overline{D}) + (A  B  \overline{C}) + (B  C  \overline{A}) + (A  \overline{C}  \overline{D}) + (A  C  D  \overline{B})\\
(B  C  D) + (B  \overline{C}  \overline{D}) + (A  C  \overline{B}  \overline{D}) + (A  D  \overline{B}  \overline{C})\\
(B  C  \overline{A}) + (B  D  \overline{C}) + (A  C  D  \overline{B}) + (A  \overline{B}  \overline{C}  \overline{D})\\
(A  \overline{D}) + (B  \overline{A}) + (B  \overline{D}) + (C  \overline{A}) + (C  \overline{D}) + (D  \overline{A}) + (A  \overline{B}  \overline{C}) + (D  \overline{B}  \overline{C})\\
(A  \overline{B}) + (A  D  \overline{C}) + (B  C  \overline{A}) + (\overline{B}  \overline{C}  \overline{D})\\
(A  \overline{D}) + (A  C  \overline{B}) + (B  C  \overline{A}) + (B  D  \overline{A})\\
(A  C  \overline{B}) + (A  C  \overline{D}) + (A  \overline{B}  \overline{D}) + (B  C  D  \overline{A})\\
(B  \overline{D}) + (C  \overline{D}) + (D  \overline{A}) + (D  \overline{B}  \overline{C})\\
(B  \overline{C}) + (B  \overline{D}) + (C  \overline{B}) + (C  \overline{D}) + (D  \overline{A}) + (D  \overline{B}) + (D  \overline{C})\\
(A  \overline{B}) + (A  \overline{D}) + (B  \overline{A}) + (B  \overline{D}) + (C  \overline{A}) + (D  \overline{A}) + (D  \overline{B})\\
(A  \overline{D}) + (A  B  \overline{C}) + (A  C  \overline{B}) + (B  C  \overline{A}) + (\overline{B}  \overline{C}  \overline{D})\\
A + B + D + \overline{C}\\
(A  C  \overline{B}) + (B  C  D  \overline{A})\\
(A  B  \overline{C}) + (A  C  \overline{B}) + (A  \overline{B}  \overline{D})\\
(A  B) + (C  D  \overline{B}) + (A  \overline{C}  \overline{D}) + (D  \overline{A}  \overline{C})\\
(A  \overline{D}) + (B  C  \overline{A}) + (B  D  \overline{A}) + (C  D  \overline{B})\\
(A  B  \overline{D}) + (A  C  \overline{B}) + (B  C  D  \overline{A})\\
(B  C  \overline{A}) + (C  D  \overline{A}) + (A  \overline{B}  \overline{D}) + (A  \overline{C}  \overline{D})\\
(B  \overline{D}) + (A  B  \overline{C}) + (B  C  \overline{A}) + (A  \overline{C}  \overline{D}) + (A  C  D  \overline{B}) + (D  \overline{A}  \overline{B}  \overline{C})\\
\overline{D} + (A  \overline{B}) + (A  \overline{C}) + (B  \overline{A}) + (B  \overline{C}) + (C  \overline{A}) + (C  \overline{B})\\
(A  C  \overline{B}) + (A  D  \overline{C}) + (B  C  D  \overline{A})\\
(A  B  \overline{C}) + (A  B  \overline{D}) + (B  C  \overline{A}) + (B  C  \overline{D}) + (B  D  \overline{A}) + (B  D  \overline{C}) + (C  D  \overline{B})\\
(A  \overline{C}  \overline{D}) + (B  C  D  \overline{A})\\
(A  D  \overline{B}) + (B  C  D  \overline{A})\\
(A  \overline{B}  \overline{C}) + (A  \overline{C}  \overline{D}) + (B  C  \overline{A}  \overline{D})\\
(B  \overline{A}  \overline{C}) + (C  \overline{B}  \overline{D}) + (D  \overline{A}  \overline{B})\\
(A  \overline{D}) + (D  \overline{A}) + (B  C  \overline{A}) + (C  D  \overline{B}) + (\overline{A}  \overline{B}  \overline{C})\\
(A  \overline{D}) + (D  \overline{A}) + (D  \overline{C}) + (\overline{B}  \overline{C}) + (B  C  \overline{A})\\
(B  \overline{A}) + (B  \overline{D}) + (A  \overline{B}  \overline{C}) + (C  \overline{A}  \overline{D}) + (D  \overline{B}  \overline{C})\\
(A  B  \overline{C}) + (A  B  \overline{D}) + (A  \overline{C}  \overline{D}) + (A  C  D  \overline{B}) + (B  C  D  \overline{A})\\
(C  \overline{D}) + (D  \overline{B}) + (D  \overline{C}) + (A  B  \overline{C})\\
(C  \overline{B}) + (D  \overline{A}) + (\overline{B}  \overline{D}) + (A  B  \overline{C})\\
(A  C  \overline{B}) + (A  C  \overline{D}) + (B  C  D  \overline{A})\\
(B  \overline{D}) + (C  \overline{A}) + (C  \overline{B}) + (C  \overline{D}) + (\overline{A}  \overline{D}) + (A  B  \overline{C}) + (A  D  \overline{B}) + (A  D  \overline{C})\\
(A  C  \overline{B}) + (A  \overline{C}  \overline{D}) + (B  C  \overline{A}  \overline{D})\\
(A  B  D  \overline{C}) + (A  C  D  \overline{B}) + (A  \overline{B}  \overline{C}  \overline{D}) + (D  \overline{A}  \overline{B}  \overline{C})\\
(A  \overline{D}) + (B  C  D) + (B  D  \overline{A}) + (C  \overline{B}  \overline{D})\\
(A  \overline{D}) + (A  B  \overline{C}) + (B  C  D  \overline{A})\\
(B  D  \overline{C}) + (A  \overline{B}  \overline{D})\\
(A  \overline{B}) + (A  \overline{C}) + (B  C  \overline{A}) + (\overline{B}  \overline{C}  \overline{D})\\
(B  \overline{A}) + (B  \overline{C}) + (C  \overline{D}) + (A  C  \overline{B}) + (D  \overline{A}  \overline{C})\\
(A  B  D  \overline{C}) + (A  C  D  \overline{B}) + (B  C  D  \overline{A})\\
(A  B  \overline{D}) + (C  \overline{B}  \overline{D})\\
(D  \overline{A}) + (D  \overline{B}) + (D  \overline{C}) + (A  B  \overline{C}) + (A  C  \overline{B}) + (B  C  \overline{A})\\
(C  \overline{B}  \overline{D}) + (A  B  \overline{C}  \overline{D})\\
(B  \overline{C}) + (D  \overline{A}) + (D  \overline{B}) + (A  B  \overline{D}) + (C  \overline{A}  \overline{B})\\
(B  \overline{C}) + (A  B  \overline{D}) + (A  C  \overline{B}) + (A  C  \overline{D}) + (A  D  \overline{B}) + (A  D  \overline{C}) + (B  D  \overline{A}) + (C  D  \overline{A}) + (C  D  \overline{B}) + (\overline{A}  \overline{C}  \overline{D})\\
(D  \overline{A}) + (D  \overline{B}) + (D  \overline{C}) + (A  C  \overline{D})\\
(A  B  \overline{C}) + (A  B  \overline{D}) + (C  D  \overline{B}) + (A  \overline{C}  \overline{D}) + (D  \overline{A}  \overline{B})\\
(B  D) + (A  B  \overline{C}) + (A  C  \overline{B}) + (B  C  \overline{A})\\
(A  B) + (A  C) + (A  \overline{D}) + (B  C  \overline{D})\\
(C  \overline{D}) + (D  \overline{C}) + (A  B  \overline{C}) + (A  D  \overline{B})\\
(C  \overline{B}) + (D  \overline{A}) + (\overline{A}  \overline{B}) + (A  \overline{C}  \overline{D})\\
(A  B  \overline{C}) + (A  C  \overline{D}) + (A  D  \overline{C}) + (B  D  \overline{A})\\
(A  \overline{B}) + (A  \overline{D}) + (B  \overline{A}) + (B  \overline{D}) + (D  \overline{A}) + (D  \overline{B})\\
(D  \overline{A}) + (D  \overline{C}) + (A  C  \overline{B})\\
(A  \overline{D}) + (B  \overline{A}) + (B  \overline{C}) + (C  \overline{D}) + (A  C  \overline{B})\\
(B  C  D) + (A  B  \overline{C}) + (A  C  \overline{B})\\
(B  C  \overline{A}) + (A  \overline{B}  \overline{D}) + (A  \overline{C}  \overline{D})\\
(A  B  \overline{C}) + (A  B  \overline{D}) + (C  D  \overline{A}) + (C  D  \overline{B})\\
(A  C  \overline{B}) + (C  \overline{B}  \overline{D}) + (A  B  \overline{C}  \overline{D})\\
(B  \overline{A}) + (C  \overline{A}) + (B  \overline{C}  \overline{D}) + (C  \overline{B}  \overline{D}) + (A  D  \overline{B}  \overline{C})\\
(A  \overline{B}) + (A  \overline{C}) + (B  \overline{A}) + (B  \overline{C}) + (C  \overline{A}) + (C  \overline{B}) + (D  \overline{A})\\
(A  \overline{D}) + (D  \overline{A}) + (D  \overline{B}) + (\overline{B}  \overline{C}) + (B  C  \overline{A})\\
(A  B  D) + (B  C  D) + (A  \overline{B}  \overline{C}  \overline{D})\\
(A  B  D) + (A  C  D) + (B  C  D) + (\overline{B}  \overline{C}  \overline{D})\\
(A  B  C  \overline{D}) + (A  C  D  \overline{B}) + (B  C  D  \overline{A}) + (A  \overline{B}  \overline{C}  \overline{D}) + (D  \overline{A}  \overline{B}  \overline{C})\\
(A  B  \overline{C}) + (A  C  \overline{B}) + (A  \overline{B}  \overline{D}) + (B  C  D  \overline{A})\\
(A  B  \overline{C}) + (A  B  \overline{D}) + (C  D  \overline{B}) + (A  \overline{C}  \overline{D})\\
A  B  C  D\\
(A  C  \overline{B}) + (B  C  \overline{A}) + (A  \overline{C}  \overline{D}) + (B  \overline{C}  \overline{D}) + (C  \overline{A}  \overline{D})\\
(A  B  D  \overline{C}) + (A  C  D  \overline{B})\\
(A  \overline{B}) + (A  \overline{C}) + (\overline{B}  \overline{C}) + (B  D  \overline{A})\\
(A  \overline{C}) + (B  \overline{D}) + (D  \overline{B}) + (D  \overline{C})\\
(A  \overline{B}) + (B  \overline{D})\\
(A  B  \overline{C}) + (A  C  \overline{D}) + (A  D  \overline{C}) + (B  D  \overline{A}) + (C  D  \overline{A})\\
(A  \overline{C}  \overline{D}) + (A  C  D  \overline{B}) + (B  C  D  \overline{A})\\
A  C  D  \overline{B}\\
(B  \overline{A}) + (C  \overline{A}) + (D  \overline{B}) + (A  \overline{B}  \overline{C})\\
(A  C  \overline{B}) + (A  \overline{C}  \overline{D}) + (B  C  D  \overline{A})\\
A + D + (B  C) + (\overline{B}  \overline{C})\\
(A  \overline{B}) + (A  \overline{C}) + (\overline{C}  \overline{D}) + (B  \overline{A}  \overline{D})\\
(D  \overline{A}) + (D  \overline{B}) + (D  \overline{C}) + (A  B  \overline{C}) + (A  B  \overline{D}) + (A  C  \overline{B}) + (A  C  \overline{D})\\
A + D + (B  C)\\
(A  \overline{D}) + (B  \overline{A}) + (D  \overline{A})\\
(A  \overline{C}  \overline{D}) + (B  D  \overline{A}  \overline{C})\\
(B  C  \overline{D}) + (B  D  \overline{A}) + (B  D  \overline{C}) + (C  D  \overline{B}) + (A  \overline{C}  \overline{D})\\
(A  \overline{B}) + (A  \overline{C}) + (A  \overline{D}) + (B  C  \overline{A})\\
(A  \overline{D}) + (C  \overline{A}) + (D  \overline{A}) + (D  \overline{B})\\
A  B  D\\
B + (A  D) + (A  \overline{C})\\
A  C  \overline{B}  \overline{D}\\
(A  B  \overline{C}) + (A  D  \overline{B}) + (B  C  \overline{A}  \overline{D})\\
B  C  D  \overline{A}\\
(A  \overline{D}) + (A  C  \overline{B}) + (B  C  \overline{A}) + (B  D  \overline{C}) + (D  \overline{A}  \overline{C})\\
(B  \overline{A}  \overline{C}) + (B  \overline{C}  \overline{D}) + (C  \overline{A}  \overline{D})\\
(A  \overline{D}) + (D  \overline{A}  \overline{B}) + (D  \overline{A}  \overline{C})\\
(A  \overline{C}  \overline{D}) + (A  C  D  \overline{B}) + (B  D  \overline{A}  \overline{C})\\
(A  \overline{D}) + (A  C  \overline{B}) + (B  C  D  \overline{A})\\
(D  \overline{A}) + (D  \overline{C}) + (A  C  \overline{B}) + (B  \overline{A}  \overline{C})\\
A  \overline{D}\\
(A  B  \overline{D}) + (B  C  D  \overline{A})\\
A + (B  C) + (B  D) + (\overline{B}  \overline{C}  \overline{D})\\
(B  D  \overline{A}) + (A  D  \overline{B}  \overline{C})\\
(A  C  \overline{D}) + (B  D  \overline{C}) + (A  \overline{B}  \overline{D})\\
(A  B  C  \overline{D}) + (A  \overline{B}  \overline{C}  \overline{D})\\
(A  C  \overline{B}) + (A  D  \overline{C}) + (B  D  \overline{A}) + (C  D  \overline{A})\\
(A  B  \overline{C}) + (A  B  \overline{D}) + (C  D  \overline{A}) + (C  D  \overline{B}) + (A  \overline{C}  \overline{D})\\
(A  B  \overline{C}  \overline{D}) + (B  C  \overline{A}  \overline{D})\\
(A  C  \overline{B}) + (B  C  \overline{A}) + (B  D  \overline{A}) + (A  \overline{C}  \overline{D})\\
(A  C) + (C  D) + (A  \overline{D}) + (B  \overline{A}) + (D  \overline{A}) + (\overline{A}  \overline{C}) + (\overline{C}  \overline{D})\\
(A  B  \overline{D}) + (A  \overline{C}  \overline{D})\\
(A  \overline{D}) + (D  \overline{A}) + (C  D  \overline{B})\\
(B  \overline{A}  \overline{C}) + (C  \overline{B}  \overline{D}) + (D  \overline{A}  \overline{C})\\
(A  \overline{D}) + (C  D  \overline{A}) + (C  D  \overline{B}) + (D  \overline{A}  \overline{B})\\
(A  B) + (A  C) + (A  \overline{D}) + (B  \overline{D}) + (D  \overline{A}  \overline{B})\\
(A  \overline{B}) + (A  \overline{C}) + (B  \overline{D}) + (\overline{B}  \overline{C}) + (B  C  \overline{A})\\
(A  \overline{D}) + (B  C  \overline{A}) + (D  \overline{A}  \overline{B})\\
(A  B  \overline{D}) + (A  \overline{C}  \overline{D}) + (A  C  D  \overline{B}) + (B  C  D  \overline{A}) + (D  \overline{A}  \overline{B}  \overline{C})\\
(A  B  \overline{C}) + (A  B  \overline{D}) + (A  C  \overline{B}) + (B  D  \overline{A})\\
\overline{C} + \overline{D} + (A  \overline{B}) + (B  \overline{A})\\
(A  B  \overline{D}) + (A  C  D  \overline{B}) + (B  C  D  \overline{A})\\
(A  \overline{B}) + (B  \overline{A}) + (B  \overline{C}) + (D  \overline{A})\\
(A  B  \overline{C}) + (A  C  \overline{B}) + (A  \overline{B}  \overline{D}) + (B  C  \overline{A}  \overline{D})\\
(B  \overline{A}) + (B  \overline{C}) + (B  \overline{D}) + (D  \overline{A}) + (D  \overline{B}) + (D  \overline{C}) + (A  C  \overline{B}) + (A  C  \overline{D})\\
(B  \overline{A}) + (C  \overline{D}) + (D  \overline{A})\\
(A  \overline{D}) + (A  B  \overline{C}) + (A  C  \overline{B}) + (B  C  \overline{D})\\
(A  \overline{B}) + (A  \overline{D}) + (B  C  \overline{D})\\
(B  \overline{C}) + (A  C  \overline{D}) + (C  D  \overline{A}) + (C  D  \overline{B})\\
(B  C  D  \overline{A}) + (A  D  \overline{B}  \overline{C})\\
(A  C  \overline{B}  \overline{D}) + (B  C  \overline{A}  \overline{D})\\
(A  D) + (B  D) + (B  C  \overline{A}) + (\overline{B}  \overline{C}  \overline{D})\\
(A  \overline{B}) + (A  \overline{C}) + (A  \overline{D}) + (B  C  \overline{A}) + (B  C  \overline{D}) + (B  D  \overline{A}) + (B  D  \overline{C}) + (C  D  \overline{A}) + (C  D  \overline{B})\\
(C  D  \overline{B}) + (A  \overline{C}  \overline{D}) + (D  \overline{A}  \overline{C})\\
(A  \overline{D}) + (D  \overline{A}) + (D  \overline{B}) + (B  C  \overline{A})\\
(A  B) + (A  C  D) + (B  C  D) + (\overline{B}  \overline{C}  \overline{D})\\
(A  \overline{D}) + (B  D  \overline{A}  \overline{C}) + (C  D  \overline{A}  \overline{B})\\
\overline{A} + (B  \overline{C}) + (C  \overline{B}) + (\overline{B}  \overline{D})\\
(A  B  \overline{C}) + (A  C  \overline{B}) + (\overline{A}  \overline{B}  \overline{C}  \overline{D})\\
(B  C  D) + (A  \overline{B}  \overline{D}) + (A  \overline{C}  \overline{D}) + (B  \overline{A}  \overline{D}) + (B  \overline{C}  \overline{D}) + (C  \overline{A}  \overline{D}) + (C  \overline{B}  \overline{D}) + (D  \overline{B}  \overline{C})\\
(A  \overline{B}) + (A  \overline{C}) + (B  \overline{D}) + (B  C  \overline{A})\\
(A  C  \overline{B}) + (C  D  \overline{A}) + (A  \overline{C}  \overline{D}) + (D  \overline{A}  \overline{B})\\
(A  \overline{C}) + (A  D  \overline{B}) + (B  C  \overline{D}) + (B  D  \overline{C})\\
(D  \overline{C}) + (\overline{B}  \overline{C}) + (A  D  \overline{B}) + (B  D  \overline{A})\\
(A  \overline{B}) + (A  D  \overline{C}) + (B  C  D  \overline{A}) + (\overline{B}  \overline{C}  \overline{D})\\
(A  \overline{D}) + (B  C  \overline{A}) + (B  D  \overline{C}) + (C  D  \overline{B})\\
(A  \overline{B}) + (A  \overline{C}) + (B  \overline{D}) + (\overline{C}  \overline{D}) + (B  C  \overline{A})\\
(A  \overline{D}) + (C  \overline{B}  \overline{D}) + (B  C  D  \overline{A})\\
\overline{C} + (A  \overline{B}) + (A  \overline{D}) + (B  \overline{A}) + (B  \overline{D}) + (D  \overline{A}) + (D  \overline{B})\\
(A  \overline{D}) + (B  \overline{A}) + (B  \overline{C}) + (D  \overline{A})\\
(B  \overline{C}) + (D  \overline{C}) + (A  B  \overline{D}) + (A  D  \overline{B}) + (B  D  \overline{A}) + (C  \overline{A}  \overline{B}  \overline{D})\\
B + C + \overline{A} + \overline{D}\\
(A  B  \overline{C}) + (B  C  \overline{A}) + (B  C  \overline{D}) + (A  C  D  \overline{B}) + (\overline{B}  \overline{C}  \overline{D})\\
(D  \overline{A}) + (A  B  \overline{C}) + (A  \overline{B}  \overline{D})\\
A + B + C + D\\
(A  \overline{B}) + (A  \overline{C}  \overline{D}) + (B  C  \overline{A}  \overline{D})\\
(A  C  \overline{D}) + (B  D  \overline{A}) + (C  D  \overline{A})\\
(D  \overline{B}) + (C  D  \overline{A}) + (A  \overline{C}  \overline{D})\\
(A  \overline{B}) + (B  \overline{A}) + (B  C  \overline{D}) + (C  D  \overline{A})\\
\overline{A} + \overline{D} + (B  \overline{C})\\
(A  B  \overline{C}) + (A  C  \overline{B}) + (A  D  \overline{B}) + (B  C  \overline{A}  \overline{D})\\
A + C + D + \overline{B}\\
(D  \overline{A}) + (D  \overline{B}) + (A  \overline{C}  \overline{D}) + (C  \overline{A}  \overline{B})\\
(C  D) + (A  \overline{D}) + (C  \overline{B}) + (D  \overline{A})\\
(A  \overline{D}) + (B  D  \overline{C}) + (C  D  \overline{B}) + (\overline{B}  \overline{C}  \overline{D})\\
(D  \overline{A}) + (D  \overline{C}) + (A  B  \overline{C}) + (A  C  \overline{D}) + (B  C  \overline{A})\\
(A  \overline{C}) + (B  \overline{D}) + (C  \overline{A}) + (D  \overline{A}) + (D  \overline{B})\\
(A  B  D  \overline{C}) + (B  C  D  \overline{A}) + (A  C  \overline{B}  \overline{D})\\
(A  B  \overline{C}) + (B  C  \overline{A}) + (B  C  \overline{D}) + (A  \overline{C}  \overline{D}) + (A  C  D  \overline{B})\\
(A  B  D  \overline{C}) + (B  C  D  \overline{A})\\
(B  \overline{A}) + (B  \overline{C}) + (A  C  \overline{B}) + (A  \overline{B}  \overline{D})\\
(A  \overline{C}) + (B  \overline{D}) + (D  \overline{A}) + (D  \overline{B})\\
(A  \overline{D}) + (B  C  D  \overline{A})\\
(A  \overline{D}) + (B  \overline{A}) + (D  \overline{A}) + (C  D  \overline{B})\\
(B  \overline{A}) + (B  \overline{C}) + (D  \overline{C}) + (A  C  \overline{B}) + (C  \overline{B}  \overline{D})\\
(A  C  D  \overline{B}) + (A  B  \overline{C}  \overline{D}) + (B  C  \overline{A}  \overline{D})\\
(B  \overline{A}) + (C  \overline{A}) + (C  D  \overline{B})\\
(B  \overline{C}) + (D  \overline{A}) + (D  \overline{B}) + (D  \overline{C}) + (\overline{A}  \overline{C}) + (A  B  \overline{D}) + (A  C  \overline{B}) + (A  C  \overline{D})\\
(A  \overline{B}  \overline{D}) + (A  \overline{C}  \overline{D}) + (B  C  D  \overline{A})\\
(A  B  C  \overline{D}) + (A  B  D  \overline{C}) + (A  C  D  \overline{B}) + (B  C  D  \overline{A})\\
(B  \overline{A}) + (B  \overline{D}) + (C  \overline{A}) + (C  \overline{D}) + (A  D  \overline{B}  \overline{C})\\
A  B  C  \overline{D}\\
A + B + (\overline{C}  \overline{D})\\
(A  B  \overline{C}) + (A  B  \overline{D}) + (A  C  \overline{B}) + (B  C  \overline{A}) + (B  C  \overline{D}) + (B  D  \overline{A}) + (B  D  \overline{C})\\
(B  C  D) + (B  \overline{C}  \overline{D}) + (C  \overline{A}  \overline{D}) + (C  \overline{B}  \overline{D}) + (D  \overline{B}  \overline{C})\\
(A  B  \overline{C}) + (A  C  \overline{B}) + (A  D  \overline{B}) + (B  C  \overline{D})\\
(C  \overline{B}  \overline{D}) + (D  \overline{A}  \overline{B}) + (B  \overline{A}  \overline{C}  \overline{D})\\
(D  \overline{A}) + (D  \overline{B}) + (A  B  \overline{C}) + (A  C  \overline{B})\\
(A  D  \overline{C}) + (B  C  D  \overline{A}) + (A  C  \overline{B}  \overline{D})\\
(D  \overline{B}) + (D  \overline{C}) + (B  C  \overline{A}) + (A  \overline{B}  \overline{C})\\
(A  C  \overline{B}) + (C  \overline{B}  \overline{D}) + (B  C  D  \overline{A}) + (A  B  \overline{C}  \overline{D})\\
(B  \overline{A}) + (A  D  \overline{B}) + (B  \overline{C}  \overline{D}) + (C  \overline{A}  \overline{D})\\
\overline{A} + \overline{B} + \overline{C} + \overline{D}\\
(B  D) + (A  \overline{B}) + (B  C  \overline{A}) + (\overline{B}  \overline{C}  \overline{D})\\
(A  B  D  \overline{C}) + (A  \overline{B}  \overline{C}  \overline{D})\\
(A  \overline{B}) + (A  \overline{C}) + (A  \overline{D}) + (D  \overline{A}) + (D  \overline{B}) + (D  \overline{C}) + (B  C  \overline{A}) + (B  C  \overline{D})\\
(A  \overline{D}) + (B  D  \overline{C}) + (C  D  \overline{B}) + (\overline{A}  \overline{B}  \overline{C})\\
(A  C  \overline{B}) + (A  \overline{C}  \overline{D}) + (D  \overline{A}  \overline{B}  \overline{C})\\
(A  B  \overline{D}) + (A  D  \overline{B}) + (A  D  \overline{C}) + (B  C  D  \overline{A})\\
(B  \overline{C}) + (C  \overline{B}) + (D  \overline{A}) + (\overline{B}  \overline{D})\\
(A  B  \overline{C}  \overline{D}) + (A  C  \overline{B}  \overline{D}) + (B  C  \overline{A}  \overline{D})\\
(A  D) + (B  D) + (A  B  \overline{C}) + (A  C  \overline{B}) + (B  C  \overline{A}) + (\overline{A}  \overline{B}  \overline{C}  \overline{D})\\
(A  B  \overline{C}) + (A  C  \overline{B}) + (B  C  \overline{A}) + (A  \overline{B}  \overline{D})\\
C + \overline{A} + \overline{D}\\
(A  \overline{D}) + (A  C  \overline{B}) + (B  C  \overline{A}) + (B  D  \overline{C})\\
(A  \overline{C}) + (A  \overline{D}) + (C  \overline{A}) + (C  \overline{D}) + (D  \overline{A}) + (D  \overline{C}) + (\overline{A}  \overline{B})\\
(A  B  \overline{C}) + (A  B  \overline{D}) + (A  \overline{C}  \overline{D})\\
(A  D  \overline{B}) + (A  \overline{C}  \overline{D}) + (B  D  \overline{A}  \overline{C})\\
(A  D) + (B  C) + (B  D) + (A  \overline{C})\\
(A  \overline{B}) + (B  \overline{D}) + (B  \overline{A}  \overline{C})\\
(D  \overline{A}) + (A  B  \overline{C}) + (A  \overline{B}  \overline{D}) + (\overline{A}  \overline{B}  \overline{C})\\
(A  B  \overline{C}) + (A  C  \overline{D}) + (A  D  \overline{C}) + (B  D  \overline{C})\\
(A  B  \overline{C}) + (A  C  \overline{B}) + (B  C  \overline{A}) + (B  D  \overline{C}) + (\overline{B}  \overline{C}  \overline{D})\\
(B  \overline{A}) + (B  \overline{C}) + (B  \overline{D}) + (D  \overline{A}) + (D  \overline{B}) + (D  \overline{C}) + (\overline{A}  \overline{C}) + (A  C  \overline{B}) + (A  C  \overline{D})\\
(A  B  D  \overline{C}) + (A  C  \overline{B}  \overline{D})\\
(A  \overline{D}) + (A  B  \overline{C}) + (C  D  \overline{A})\\
(A  \overline{D}) + (D  \overline{A}) + (B  C  \overline{A})\\
(A  B  \overline{D}) + (A  C  \overline{B}) + (B  D  \overline{A}) + (C  D  \overline{B})\\
(B  D) + (A  \overline{B}) + (A  \overline{C}) + (B  C  \overline{A}) + (\overline{B}  \overline{C}  \overline{D})\\
(A  \overline{D}) + (A  B  \overline{C}) + (A  C  \overline{B}) + (B  C  \overline{A}) + (\overline{A}  \overline{B}  \overline{C})\\
\overline{B} + (A  \overline{C}) + (A  \overline{D}) + (C  \overline{A}) + (C  \overline{D}) + (D  \overline{A}) + (D  \overline{C})\\
(B  C  D  \overline{A}) + (A  C  \overline{B}  \overline{D}) + (A  D  \overline{B}  \overline{C})\\
(A  B  \overline{C}) + (A  C  \overline{B}) + (B  C  D  \overline{A})\\
(B  D) + (A  C  D) + (B  C  \overline{A}) + (A  \overline{C}  \overline{D})\\
(A  \overline{D}) + (D  \overline{B}) + (B  C  \overline{D}) + (D  \overline{A}  \overline{C})\\
(A  C  \overline{B}) + (C  \overline{A}  \overline{D}) + (A  B  \overline{C}  \overline{D})\\
(A  B  \overline{C}) + (A  C  \overline{B}) + (B  C  \overline{A})\\
(A  \overline{B}) + (A  \overline{C}) + (A  \overline{D}) + (B  \overline{A}) + (B  \overline{C}) + (B  \overline{D}) + (C  \overline{A}) + (C  \overline{B}) + (C  \overline{D}) + (D  \overline{A}) + (D  \overline{B}) + (D  \overline{C})\\
(A  \overline{D}) + (C  D  \overline{B}) + (D  \overline{A}  \overline{C})\\
(A  B  \overline{C}) + (A  B  \overline{D}) + (A  \overline{C}  \overline{D}) + (A  C  D  \overline{B})\\
(C  \overline{A}) + (D  \overline{A}) + (B  \overline{C}  \overline{D}) + (D  \overline{B}  \overline{C})\\
(A  \overline{B}) + (A  \overline{C}) + (A  \overline{D}) + (B  D  \overline{A}) + (B  D  \overline{C}) + (C  D  \overline{A}) + (C  D  \overline{B}) + (\overline{B}  \overline{C}  \overline{D})\\
(B  C  D) + (B  \overline{C}  \overline{D}) + (A  C  \overline{B}  \overline{D})\\
A  \overline{C}  \overline{D}\\
(B  D  \overline{A}) + (B  D  \overline{C}) + (A  C  D  \overline{B}) + (\overline{A}  \overline{B}  \overline{C}) + (\overline{B}  \overline{C}  \overline{D})\\
(A  \overline{D}) + (B  C  \overline{D})\\
(A  C  \overline{B}) + (B  C  \overline{A}) + (C  \overline{A}  \overline{D}) + (A  B  \overline{C}  \overline{D})\\
(D  \overline{A}) + (A  B  \overline{D}) + (A  C  \overline{D}) + (B  C  \overline{A})\\
(A  \overline{D}) + (B  C  D) + (C  \overline{B}  \overline{D})\\
(A  C  D  \overline{B}) + (B  C  D  \overline{A})\\
A  D  \overline{B}  \overline{C}\\
(D  \overline{A}) + (B  C  \overline{A}) + (A  \overline{B}  \overline{D}) + (A  \overline{C}  \overline{D})\\
(A  B  \overline{C}) + (A  B  \overline{D}) + (A  \overline{C}  \overline{D}) + (B  \overline{C}  \overline{D}) + (A  C  D  \overline{B}) + (B  C  D  \overline{A})\\
\overline{A} + (B  \overline{D}) + (C  \overline{D}) + (D  \overline{B}  \overline{C})\\
(B  \overline{A}) + (B  \overline{C}  \overline{D}) + (C  \overline{B}  \overline{D}) + (A  D  \overline{B}  \overline{C})\\
A + B + C + \overline{D}\\
(A  \overline{D}) + (A  B  \overline{C}) + (A  C  \overline{B})\\
(A  \overline{B}) + (B  C  \overline{D}) + (B  D  \overline{A}) + (C  D  \overline{A})\\
(A  B) + (A  C  D) + (B  C  \overline{D}) + (B  D  \overline{C}) + (C  D  \overline{B}) + (A  \overline{C}  \overline{D}) + (D  \overline{A}  \overline{B}) + (D  \overline{A}  \overline{C})\\
(A  \overline{D}) + (B  \overline{C}) + (C  \overline{B}) + (D  \overline{A}) + (\overline{A}  \overline{B})\\
(A  \overline{D}) + (B  \overline{C}) + (D  \overline{A}) + (C  D  \overline{B})\\
(A  C  \overline{B}) + (C  D  \overline{B}) + (A  \overline{C}  \overline{D}) + (D  \overline{A}  \overline{C})\\
(A  \overline{B}) + (B  C  \overline{D}) + (B  D  \overline{A})\\
(A  \overline{B}) + (C  \overline{A}) + (B  \overline{C}  \overline{D})\\
(A  \overline{D}) + (B  \overline{A}) + (B  \overline{C}) + (\overline{A}  \overline{C}) + (A  C  \overline{B})\\
(A  C) + (A  \overline{D}) + (B  \overline{A}) + (C  \overline{D})\\
(A  B) + (C  D) + (A  \overline{D}) + (B  \overline{D}) + (D  \overline{A}  \overline{B})\\
(A  C  \overline{B}) + (A  B  D  \overline{C}) + (B  C  D  \overline{A}) + (\overline{A}  \overline{B}  \overline{C}  \overline{D})\\
(A  \overline{D}) + (B  D  \overline{A}  \overline{C})\\
(B  \overline{A}) + (B  \overline{C}) + (C  \overline{D}) + (D  \overline{C}) + (A  D  \overline{B})\\
A + (B  C  D) + (\overline{B}  \overline{C}  \overline{D})\\
(A  \overline{D}) + (B  \overline{D}) + (D  \overline{A}  \overline{B}) + (D  \overline{A}  \overline{C})\\
(A  C  \overline{D}) + (B  D  \overline{C}) + (A  \overline{B}  \overline{D}) + (D  \overline{A}  \overline{C})\\
(C  D  \overline{A}  \overline{B}) + (A  \overline{B}  \overline{C}  \overline{D})\\
(A  \overline{C}) + (D  \overline{A}) + (D  \overline{B}) + (B  C  \overline{D})\\
(B  \overline{D}) + (C  \overline{A}) + (D  \overline{A}) + (D  \overline{B}  \overline{C})\\
(B  \overline{C}) + (B  \overline{D}) + (D  \overline{A}) + (A  C  \overline{B})\\
(A  B  \overline{C}) + (A  C  \overline{B}) + (A  D  \overline{B}) + (B  C  \overline{A}) + (\overline{A}  \overline{B}  \overline{D})\\
(B  \overline{D}) + (C  \overline{A}) + (D  \overline{A}) + (A  \overline{B}  \overline{C})\\
(A  \overline{B}) + (B  \overline{A}) + (B  \overline{C}) + (\overline{C}  \overline{D})\\
(A  B) + (A  C) + (A  \overline{D}) + (B  C  \overline{D}) + (D  \overline{A}  \overline{B}  \overline{C})\\
(A  \overline{D}) + (D  \overline{A}) + (B  C  \overline{A}) + (D  \overline{B}  \overline{C})\\
(A  C  \overline{B}) + (A  D  \overline{C}) + (B  D  \overline{A})\\
(A  \overline{B}) + (A  \overline{C}) + (D  \overline{A}) + (\overline{B}  \overline{C}) + (B  C  \overline{A})\\
(C  \overline{A}) + (D  \overline{A}) + (D  \overline{B}) + (A  \overline{C}  \overline{D})\\
(A  B  D  \overline{C}) + (B  C  D  \overline{A}) + (A  \overline{B}  \overline{C}  \overline{D})\\
B  \overline{C}  \overline{D}\\
(B  \overline{A}) + (A  D  \overline{B}) + (B  \overline{C}  \overline{D}) + (C  \overline{B}  \overline{D})\\
(A  \overline{D}) + (B  D  \overline{A}) + (C  D  \overline{B})\\
(B  \overline{A}) + (B  \overline{C}) + (D  \overline{B}) + (A  C  \overline{B})\\
(A  \overline{D}) + (D  \overline{A})\\
A  D  \overline{B}\\
(A  \overline{D}) + (D  \overline{A}) + (\overline{B}  \overline{C}) + (B  C  \overline{A})\\
(A  B  \overline{D}) + (A  C  \overline{B}) + (C  D  \overline{B}) + (B  D  \overline{A}  \overline{C})\\
(A  \overline{D}) + (D  \overline{A}) + (B  C  \overline{A}) + (B  D  \overline{C}) + (\overline{A}  \overline{B}  \overline{C})\\
(B  C  \overline{A}) + (A  B  D  \overline{C}) + (A  C  D  \overline{B})\\
\overline{A} + \overline{C} + \overline{D}\\
(A  B  \overline{D}) + (A  C  \overline{D}) + (A  D  \overline{B}  \overline{C})\\
(A  B  \overline{C}) + (B  D  \overline{C}) + (C  D  \overline{A}) + (A  \overline{B}  \overline{D})\\
(A  D  \overline{B}) + (A  D  \overline{C})\\
A  \overline{B}  \overline{C}  \overline{D}\\
(A  B  C  \overline{D}) + (A  B  D  \overline{C}) + (A  C  D  \overline{B}) + (B  C  D  \overline{A}) + (\overline{B}  \overline{C}  \overline{D})\\
(D  \overline{A}) + (A  C  \overline{D}) + (B  D  \overline{C}) + (C  D  \overline{B}) + (A  \overline{B}  \overline{D})\\
(A  C  \overline{B}) + (A  B  D  \overline{C})\\
(A  B  \overline{C}) + (A  \overline{C}  \overline{D}) + (A  C  D  \overline{B})\\
(A  C  \overline{D}) + (B  C  \overline{A}) + (B  D  \overline{A}) + (C  D  \overline{A})\\
(A  \overline{C}) + (A  \overline{D}) + (C  \overline{A}) + (C  \overline{D}) + (D  \overline{A}) + (D  \overline{B}) + (D  \overline{C})\\
(B  D  \overline{A}) + (C  D  \overline{A}) + (A  C  \overline{B}  \overline{D}) + (A  D  \overline{B}  \overline{C})\\
(A  \overline{D}) + (B  \overline{A}) + (B  \overline{C}) + (D  \overline{A}) + (\overline{A}  \overline{C})\\
A + B + D\\
(A  \overline{D}) + (C  \overline{D}) + (A  C  \overline{B}) + (B  C  \overline{A})\\
(A  B  C) + (A  B  D) + (A  C  D) + (B  C  D) + (\overline{B}  \overline{C}  \overline{D})\\
(A  B  \overline{D}) + (A  \overline{C}  \overline{D}) + (A  C  D  \overline{B}) + (B  C  D  \overline{A})\\
(A  B  \overline{C}  \overline{D}) + (B  D  \overline{A}  \overline{C})\\
(D  \overline{A}) + (B  C  \overline{D}) + (C  D  \overline{B}) + (A  \overline{B}  \overline{D})\\
(C  D  \overline{B}) + (A  \overline{C}  \overline{D})\\
(A  \overline{B}) + (A  \overline{C}) + (B  C  \overline{A}  \overline{D})\\
\overline{D} + (A  \overline{B}) + (B  \overline{A}) + (B  \overline{C})\\
(A  B  \overline{C}) + (A  C  \overline{B})\\
(A  D  \overline{C}) + (B  C  D  \overline{A})\\
(A  C  D  \overline{B}) + (A  \overline{B}  \overline{C}  \overline{D})\\
(A  C  \overline{B}) + (B  C  D  \overline{A}) + (A  B  \overline{C}  \overline{D})\\
(C  D  \overline{B}) + (A  \overline{C}  \overline{D}) + (B  D  \overline{A}  \overline{C})\\
(A  B  \overline{D}) + (A  C  \overline{B}) + (B  D  \overline{A})\\
C + \overline{A} + \overline{B} + \overline{D}\\
(A  B  \overline{C}) + (A  D  \overline{C}) + (B  D  \overline{A}) + (C  D  \overline{A})\\
\overline{A} + \overline{D} + (\overline{B}  \overline{C})\\
(A  B  \overline{D}) + (A  C  \overline{B}) + (B  D  \overline{C})\\
(A  \overline{B}) + (A  \overline{C}) + (D  \overline{C}) + (\overline{B}  \overline{C}) + (B  C  \overline{A})\\
(A  B  \overline{C}) + (A  C  \overline{B}) + (A  C  \overline{D}) + (C  D  \overline{A}) + (B  \overline{C}  \overline{D})\\
A  B  \overline{C}  \overline{D}\\
(B  D  \overline{A}) + (C  D  \overline{A}) + (A  D  \overline{B}  \overline{C})\\
(C  \overline{B}  \overline{D}) + (B  \overline{A}  \overline{C}  \overline{D}) + (D  \overline{A}  \overline{B}  \overline{C})\\
(A  B  \overline{D}) + (A  D  \overline{B}  \overline{C})\\
B + (C  D) + (A  \overline{D}) + (D  \overline{A})\\
(A  C  \overline{D}) + (A  D  \overline{B}) + (A  D  \overline{C}) + (B  C  D  \overline{A})\\
B + (A  D) + (\overline{C}  \overline{D})\\
(B  \overline{C}  \overline{D}) + (A  C  \overline{B}  \overline{D})\\
(A  B  \overline{D}) + (C  \overline{B}  \overline{D}) + (B  C  D  \overline{A})\\
(A  C  \overline{B}) + (A  B  D  \overline{C}) + (B  C  D  \overline{A})\\
(A  \overline{B}  \overline{D}) + (A  B  D  \overline{C})\\
(A  \overline{B}  \overline{D}) + (A  B  D  \overline{C}) + (B  C  D  \overline{A})\\
(A  B  C  \overline{D}) + (A  C  D  \overline{B}) + (B  C  D  \overline{A}) + (A  \overline{B}  \overline{C}  \overline{D})\\
(A  \overline{D}) + (B  C  \overline{D}) + (B  D  \overline{C}) + (C  D  \overline{B})\\
A  B  \overline{D}\\
(A  \overline{D}) + (B  \overline{A}) + (C  \overline{B}) + (D  \overline{A})\\
(A  B  \overline{D}) + (A  C  \overline{B}) + (C  \overline{B}  \overline{D}) + (B  C  D  \overline{A})\\
(B  C  \overline{A}) + (A  \overline{C}  \overline{D})\\
B + \overline{A} + \overline{C} + \overline{D}\\
(B  \overline{A}) + (C  \overline{B}  \overline{D}) + (D  \overline{A}  \overline{C})\\
(A  \overline{D}) + (B  \overline{A}) + (B  \overline{C}) + (D  \overline{A}) + (C  D  \overline{B})\\
(D  \overline{A}) + (D  \overline{C}) + (A  C  \overline{B}  \overline{D})\\
(A  B  \overline{D}) + (A  C  \overline{B})\\
(A  C  \overline{B}) + (A  \overline{B}  \overline{D}) + (A  B  D  \overline{C})\\
(A  B  \overline{D}) + (A  C  \overline{D}) + (B  D  \overline{C}) + (C  D  \overline{B})\\
(A  \overline{C}) + (C  \overline{A}) + (D  \overline{A}) + (D  \overline{B}) + (\overline{A}  \overline{B})\\
B + (A  C) + (C  D) + (A  \overline{D}) + (D  \overline{A}) + (\overline{A}  \overline{C}) + (\overline{C}  \overline{D})\\
(B  \overline{C}) + (A  C  \overline{B}) + (B  D  \overline{A}) + (A  \overline{B}  \overline{D})\\
(A  \overline{B}) + (B  \overline{D}) + (D  \overline{B}) + (D  \overline{C})\\
(B  C  \overline{A}) + (B  D  \overline{C}) + (A  C  D  \overline{B}) + (\overline{A}  \overline{B}  \overline{C}  \overline{D})\\
(A  D  \overline{B}) + (B  C  \overline{D}) + (A  \overline{C}  \overline{D}) + (B  D  \overline{A}  \overline{C})\\
(A  C  \overline{B}) + (A  C  \overline{D}) + (A  \overline{B}  \overline{D})\\
(C  \overline{A}) + (C  \overline{B}) + (C  \overline{D}) + (D  \overline{A}) + (D  \overline{B}) + (D  \overline{C}) + (A  B  \overline{C}) + (A  B  \overline{D})\\
(A  C) + (A  \overline{D}) + (B  D  \overline{A}) + (B  \overline{A}  \overline{C}) + (C  \overline{B}  \overline{D})\\
(A  B  D  \overline{C}) + (A  C  D  \overline{B}) + (B  C  D  \overline{A}) + (\overline{A}  \overline{B}  \overline{C}  \overline{D})\\
(A  \overline{B}) + (A  \overline{C}) + (D  \overline{A}) + (B  C  \overline{A})\\
(A  B  \overline{C}) + (A  C  \overline{B}) + (B  D  \overline{A})\\
(A  B  \overline{D}) + (A  D  \overline{B}) + (A  D  \overline{C}) + (B  D  \overline{A}) + (B  D  \overline{C}) + (C  D  \overline{A}) + (C  D  \overline{B})\\
A  C  \overline{B}\\
C  \overline{B}  \overline{D}\\
(C  D) + (D  \overline{A}) + (A  \overline{B}  \overline{D}) + (A  \overline{C}  \overline{D})\\
(A  \overline{D}) + (D  \overline{A}) + (C  D  \overline{B}) + (\overline{A}  \overline{B}  \overline{C})\\
(A  \overline{D}) + (D  \overline{A}) + (B  C  \overline{A}) + (C  D  \overline{B})\\
(A  \overline{B}) + (A  \overline{D}) + (B  \overline{A}) + (B  \overline{D}) + (D  \overline{A}) + (D  \overline{B}) + (D  \overline{C})\\
(A  \overline{D}) + (B  \overline{C}) + (C  \overline{A}) + (D  \overline{A})\\
(A  B  \overline{C}) + (A  C  \overline{B}) + (B  C  \overline{A}) + (\overline{A}  \overline{B}  \overline{C}  \overline{D})\\
(A  \overline{B}) + (A  \overline{C}) + (B  \overline{A}  \overline{D})\\
(A  B  \overline{C}) + (A  C  \overline{B}) + (A  D  \overline{B})\\
(A  \overline{D}) + (A  C  \overline{B}) + (C  \overline{B}  \overline{D}) + (B  C  D  \overline{A})\\
(\overline{A}  \overline{B}) + (C  D  \overline{B}) + (A  \overline{C}  \overline{D}) + (D  \overline{A}  \overline{C})\\
(B  \overline{A}) + (C  \overline{A}) + (B  \overline{C}  \overline{D}) + (A  D  \overline{B}  \overline{C})\\
(A  B  \overline{C}) + (A  C  \overline{B}) + (A  \overline{B}  \overline{D}) + (B  \overline{C}  \overline{D}) + (B  C  D  \overline{A})\\
(A  B  \overline{C}) + (A  B  \overline{D}) + (A  C  D  \overline{B}) + (B  C  D  \overline{A})\\
(D  \overline{A}) + (A  C  \overline{D}) + (B  D  \overline{C}) + (C  D  \overline{B}) + (A  \overline{B}  \overline{D}) + (B  \overline{A}  \overline{C})\\
C + (A  \overline{D}) + (D  \overline{A}) + (\overline{A}  \overline{B})\\
(A  \overline{D}) + (B  \overline{A}) + (B  \overline{D}) + (D  \overline{A}) + (\overline{A}  \overline{C}) + (\overline{C}  \overline{D}) + (A  C  \overline{B}) + (C  D  \overline{B})\\
(A  C  \overline{B}) + (A  D  \overline{C})\\
(C  \overline{D}) + (D  \overline{C}) + (A  B  \overline{C}) + (A  D  \overline{B}) + (B  D  \overline{A})\\
(A  \overline{B}) + (A  \overline{D}) + (B  \overline{A}) + (B  \overline{D}) + (D  \overline{A}) + (D  \overline{B}) + (\overline{A}  \overline{C})\\
(A  \overline{B}  \overline{C}) + (A  \overline{B}  \overline{D}) + (A  \overline{C}  \overline{D}) + (B  C  \overline{A}  \overline{D})\\
(A  C  \overline{D}) + (B  \overline{A}  \overline{C}) + (C  \overline{A}  \overline{B}) + (D  \overline{A}  \overline{B})\\
A + B + \overline{C} + \overline{D}\\
(A  D  \overline{C}) + (B  D  \overline{A}) + (C  D  \overline{A}) + (A  C  \overline{B}  \overline{D})\\
(A  B) + (A  C) + (B  C) + (B  D) + (C  D) + (A  \overline{D}) + (D  \overline{A}) + (\overline{A}  \overline{B}  \overline{C}) + (\overline{B}  \overline{C}  \overline{D})\\
(B  \overline{A}) + (D  \overline{A}) + (C  \overline{B}  \overline{D})\\
(C  D  \overline{A}) + (A  C  \overline{B}  \overline{D}) + (A  D  \overline{B}  \overline{C})\\
$


\end{document}